\documentclass[twocolumn,showpacs,preprintnumbers,amsmath,amssymb]{revtex4}

\usepackage{graphicx}
\usepackage{dcolumn}
\usepackage{bm}

\begin{document}

\title{First Principles Calculations of Fe on GaAs (100).}

\author{S. Mirbt, B. Sanyal, C. Isheden, and B. Johansson} 

\address{Department of Physics, Uppsala University, 
 Uppsala, Sweden} 

\date{\today}

\begin{abstract}
We have calculated from first principles the electronic structure of 
0.5 monolayer upto 5 monolayer thick Fe layers on top of a GaAs (100) surface. 
We find the Fe magnetic moment to be determined by the Fe-As distance.
As segregates to the top of the Fe film, whereas Ga most likely is found within the Fe film.
Moreover, we find an asymmetric in-plane contraction
of our unit-cell along with an expansion perpendicular to the surface.
We predict the number of Fe $3d$-holes to increase
with increasing Fe thickness on $p$-doped GaAs.

\end{abstract}
\vspace{20mm}
\pacs{75.70.-i,82.65.+r,75.50.Bb,72.25.Mk}

\maketitle

\section{Introduction}
In the context of magnetoelectronics it is important to understand the
structural and electronic properties of the interface between Fe and
GaAs \cite{general,zhu}. Several experimental investigations have been performed on this
interface \cite{krebs,filipe,gester,rahmoune,lallaizon,bensch,Xu,zhang,gordon,freeland,wedler,chambers,ruckman,kneedler,brockmann}. Under As-rich conditions the Fe film growing on top of GaAs
is found to have a reduced magnetic moment, or even a zero magnetic
moment close to the interface \cite{krebs,filipe,gester}. Moreover, structural investigations of the
Fe film lead to the conclusion that between a bcc Fe film and the GaAs
substrate there exists an intermediate phase, $Fe_{x}Ga_{y}As_{z}$
\cite{rahmoune,lallaizon}. Under
Ga-rich conditions no traces of an intermediate phase are found and the
magnetic moment is found to be bulk-like even close to the interface
\cite{bensch,Xu}.

Other theoretical investigations of the Fe/GaAs(100) interface
\cite{theory,hong,erwin,kosuth} have been performed. Except of the
calculations by Erwin et al \cite{erwin}, all other calculations
considered only an ideal Fe/GaAs interface. Erwin 
focused on Fe adatom growth and magnetic properties of the
interface and allowed for ionic relaxations only perpendicular to the
surface (i.e along the z-direction). In general their results agree
with our calculations.
In this paper we present an investigation of the (completely) relaxed
Fe/GaAs interface structure. We study the Fe magnetic moment as a
function of the Fe film thickness and the relaxed structure of the Fe film.
In addition we study As and Ga segregation. We discuss the origin of
the so called magnetically dead layers on top of GaAs and why Fe growth
differs on As terminated  and Ga terminated interfaces.

\section{Computational details}
We have performed self-consistent first-principles density functional
calculations employing a plane wave pseudopotential code (VASP )
\cite{vasp}. 
 PAW pseudopotentials \cite{paw} with an energy cutoff of 24.61 Ryd were
used.
Exchange correlation was treated within the generalized gradient
approximation (GGA)
\cite{gga}. We used a (4x4x4) folding of
special k-points \cite{monkhorst}. We used a unit cell having 6
semiconducting 
layers (3 Ga and 3 As) as the substrate and 1 to 10 Fe atoms (0.5
monolayer to 5 monolayers)
 on top of the GaAs substrate. 
A vacuum of 10 \AA \ thickness or more was kept to 
avoid interactions between neighboring unit cells 
in the (001) direction.
To simulate a bulk semiconductor, the lowermost Ga/As layer was
passivated with 
pseudohydrogen atoms \cite{pseudohyrdo}. Volume and shape relaxations
were allowed
along with the relaxation of the atoms. All atoms were relaxed except the
2 bulk 
semiconducting atoms.
It is worth mentioning that the bcc Fe is perfectly lattice
matched to the GaAs substrate (zinc-blende lattice). The
GaAs lattice constant (5.65 \AA\ ) is almost twice of the lattice
constant of bcc Fe (2.87 \AA\ ). Thus
the GaAs unit cell is equivalent to two bcc unit cells. 
We used a lattice constant of $5.735$ \AA\, which corresponds to the
calculated GaAs bulk
equilibrium lattice constant within GGA. The lattice constant of bulk Fe
is
calculated to be only $0.05 \%$ smaller. 
 The energies were converged with an accuracy of $10^{-4}$ eV. 
The Pulay stress we find to be negligible, being of the order of 0.1 kB.

Local magnetic moments were obtained by projecting wave functions onto
spherical harmonics 
within spheres centered on the atoms \cite{eichler}. We used $1.302$ \AA\,
$1.402$
\AA\, and $1.355$ \AA\ for Fe, Ga, and As spheres, respectively.
In order to compare energies of different interface geometries with a
non-equivalent number of atoms, we used the following definition of the
formation energy:
\begin{equation}
E_{form}=E_{total}-\sum_{i} N_{i} \mu _{i} .
\end{equation}
$N_{i}$ is the number of non equivalent atoms of species $i$, and $\mu
_{i}$ is the
chemical potential. We estimated the chemical potential by calculating 
the total energy of species $i$ in its bulk form. 

\section{GaAs surface }
The GaAs (100) surface structure depends on the growth environment.
Under As rich conditions the GaAs (100) surface shows a c(4x4)
\cite{c4x4}
reconstruction, whereas under Ga rich conditions the GaAs (100) surface
shows an $\epsilon$(4x2) \cite{epsilon} reconstruction. All these
reconstructions differ from an ideally terminated surface -where
either only Ga or only As exists at the surface- by the existence of Ga-Ga
and (or) As-As dimers. 

In a first attempt to understand the interface
between Fe and GaAs we neglect any surface reconstructions and assume an
ideal Ga(As) terminated surface.
In order to justify this approximation, let us assume we would grow Fe on top of some
reconstructed GaAs surface. The interaction of Fe with the dangling
bonds of As or Ga is already included in our calculation assuming an
ideal cut. The question is now what happens with the As-As (Ga-Ga)
dimers, when Fe is present at the surface? 
In a recent paper Erwin et al. \cite{erwin} calculate that Ga and As
surface dimers become unstable under Fe adsorption and Fe-Ga resp. Fe-As
bonds form instead.
This implies that our calculation assuming an ideal cut surface covers
most of the physics at the Fe-GaAs interface. 
Moreover, experimental results by Kneedler et al. \cite{kneedler} warrant our approximation to ignore the surface
reconstruction details. They investigated the influence of the GaAs
surface reconstructions on the properties of the growing Fe film. They
compared two As-rich terminations, the 2x4 and the c(4x4). 
In summary they find, although the Fe island morphology is
different, the growth mode, interfacial structure, magnetic behaviour,
and the uniaxial anisotropy to be independent on
the chosen As-rich surface reconstruction. 

\begin{figure}
\resizebox{6.5cm}{!}{\includegraphics[angle=270]{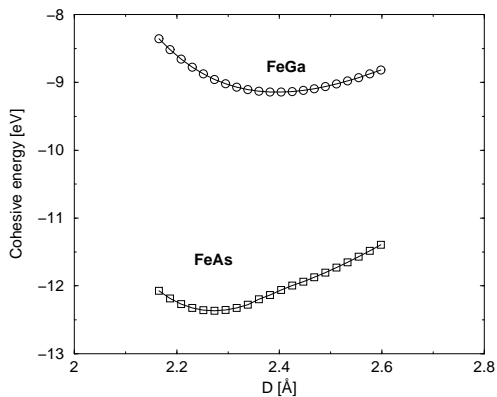}}
\caption{
Cohesive energy as a function of the Fe-X distance (X=As,Ga) for a bulk
zincblende FeX structure. The upper curve (circles) corresponds to
zincblende FeGa, the lower curve (squares) to zincblende FeAs.
}
\end{figure}
The difference in Fe growth between a Ga-terminated and an As-terminated
surface can be explained largely by kinetic arguments. In Fig.1 we show the
cohesive energy as a function of the Fe-Ga (Fe-As) distance for bulk
FeGa (FeAs) in a zincblende structure. (These artificial structures only
serve as a tool to understand the FeAs and FeGa interaction.) The cohesive energy of zincblende
FeAs is about $4$ eV lower than of FeGa. Thus the interaction between Ga
and Fe is much weaker than between As and Fe, because As has 
three $p$-electrons and Ga only one $p$-electron, which  
gives rise to a larger $pd$-hybridization between As and Fe.
According to Eq.2, energy is gained, when Fe replaces a Ga atom.
Therefore, if it is kinetically possible, Fe
replaces a Ga atom in the interface region independent of the surface
termination. 
In addition, energy is gained, -independently of the termination-, if As segregates to the surface (see section VI). The experimentally observable
physical properties of the Fe on GaAs(100) system are thus governed
by extrinsic effects like growth conditions and in addition the surface
termination governs the kinetic conditions for the Fe growth, i.e
the height of involved barriers, diffusion probability, probability
distribution of As atoms versus Ga atoms within the Fe film. 

In this paper we discuss the physical properties of an Fe film
on an As-terminated surface, but our conclusions are valid independent
of the termination, since we do not calculate any kinetic properties.

\section{Structural properties}

\begin{figure}
\resizebox{6.5cm}{!}{\includegraphics[angle=270]{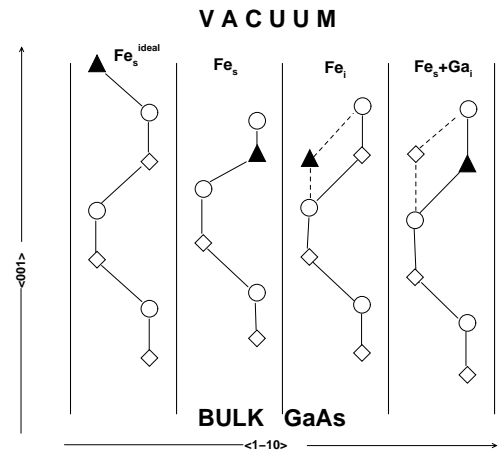}}
\caption{
Geometrical structure of 0.5ML Fe on top of GaAs(100). Circles
(diamonds)
represent As (Ga) and triangles represent Fe. The four columns
correspond to four different interface configurations as discussed in the text.
The atomic positions are shown along the $<1\bar{1}0>$ direction, bonds are
indicated by the lines connecting the atoms.}
\end{figure}

In GaAs bulk an Fe impurity is most likely found in a Ga-substitutional
position, i.e Fe replaces a Ga atom. For 0.5 ML Fe on top of a (100)
GaAs surface, we compare here four different Fe configurations (Fig.2).
We show the four different Fe
configurations in the order of decreasing formation
energy: $Fe_{s}^{ideal}$ corresponds to one Fe
atom (i.e. 0.5 ML) sitting on top of GaAs at the position of a Ga atom, that
is Fe sits substitutionally (s) following the ideal stacking (no lattice
relaxations).
$Fe_{s}$ corresponds to one Fe atom having replaced the Ga
atom closest to the surface, that is the Fe atom becomes buried under
the surface. $Fe_{i}$ corresponds to one Fe atom sitting in a
buried position but at an interstitial (i) position of the ideal GaAs
lattice. $Fe_{s} + Ga_{i}$ finally corresponds to one Fe atom
sitting in a substitutionally buried position, but in addition a Ga atom
is now sitting in an interstitial position. 

\begin{figure}
\resizebox{6.5cm}{!}{\includegraphics[angle=270]{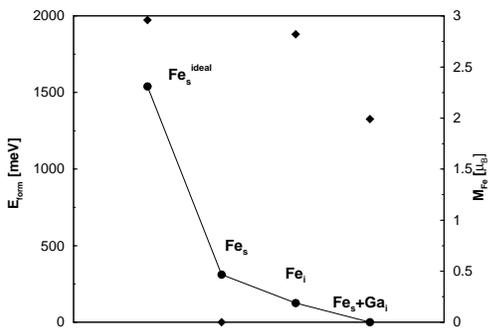}}
\caption{Formation energy, $E_{form}$, of 0.5ML Fe configurations (see Fig.2) relative to the
lowest configuration, $Fe_{s}+Ga_{i}$. Also indicated is the magnetic moment of the Fe
atom (right scale and diamonds).}
\end{figure}

In Fig.3 we show the formation energy relative to the
formation energy of the $Fe_{s} + Ga_{i}$ configuration. 
Let us now discuss these results:
The four configurations differ by the number of vacant neighbours that each
Fe atom has: In the $Fe_{s}^{ideal}$ configuration, Fe has four vacant
neighbours, in the $Fe_{s}$, Fe has three vacant neighbours, and in the
other two configurations, Fe has only two vacant neighbours.
The more neighbours the Fe atom has, the more energy is gained due to
increased wavefunction overlap. Thus the $Fe_{s}^{ideal}$ configuration
is highest in energy, because the wavefunction overlap between Fe and
its surrounding is minimal.
Next we consider the $Fe_{s}$ configuration, which is already $1$ eV
lower than $Fe_{s}^{ideal}$. Besides having one more neighbour than
$Fe_{s}^{ideal}$, in the
$Fe_{s}$  configuration two Fe-As bonds have formed. For this
configuration the Fe magnetic moment is totally quenched. This will be
further discussed in the next section.
The $Fe_{i}$ and $Fe_{s} + Ga_{i}$ configurations are lower than
$Fe_{s}$ because they only have two vacant neighbours. These two
configurations are similar
except that the topmost Ga and Fe positions have been interchanged.
We find from our calculation 
\begin{equation}
E_{Ga-As} + E_{Fe_{i}} > E_{Fe-As} + E_{Ga_{i}},
\end{equation}
where $E_{Ga-As}$ ($E_{Fe-As}$) is the formation energy of a Ga-As (Fe-As) surface bond
and $E_{Fe_{i}}$ ($E_{Ga_{i}}$) is the formation energy of an interstitial surface defect $Fe_{i} (Ga_{i})$ .

We compared the charge density and performed an (unrelaxed) calculation without the
respective interstitial atom.
In both configurations, the top Ga atom forms together with the Fe atom a metallic layer.
The difference is the bonding: For $Fe_{i}$ we have mainly two Ga-As
bonds and one Ga-Fe metallic bond, whereas for  $Fe_{s} + Ga_{i}$ we
have two Fe-As bonds and again one Ga-Fe metallic bond, but in addition
the interstitial Ga atom bonds to it's As neighbours. This $Ga_{i}$-As
interaction reduces the Fe-As interaction, whereby the Fe magnetic
moment is close to it's bulk value. It is thus the $Ga_{i}$ that lowers
the energy of the $Fe_{s} + Ga_{i}$  configuration. 
In summary, a dilute
(sub-monolayer) Fe film on top of GaAs (100) will break the top GaAs
bonds and create Fe-As bonds and Ga$_{i}$ instead. 

\begin{figure}
\resizebox{6.5cm}{!}{\includegraphics[angle=270]{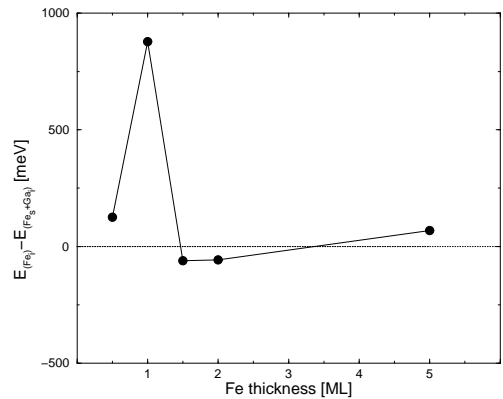}}
\caption{Energy difference between an interface configuration where Fe sits
interstitial ($Fe_{i}$) and where Fe sits substitutional together with
an interstitial Ga ($Fe_{s}+Ga_{i}$) as a function of the Fe
film thickness.}
\end{figure}

Even for thicker Fe films this is in principle still valid. But an Fe interface
atom bounded to an Fe film will have a reduced mobility compared to an
Fe adatom. This might prevent Fe from dissolving GaAs. This kinetic issue we have
not studied any further. We calculate an $Fe_{s}^{ideal}$ IC to be 
preferred if As and Ga segregation is neglected.
This result is in agreement with a calculation by Erwin et al \cite{erwin}.
But including As surface segregation (still neglecting Ga segregation) we find the $Fe_{s} + Ga_{i}$ IC to have 
the lowest energy almost independent of the Fe film thickness.  
In Fig.4 we show the energy difference between the $Fe_{i}$
and $Fe_{s} + Ga_{i}$ interface configurations (IC) for a varying number of Fe
thickness. For $0.5, 1$ and $5$ ML of Fe we calculate the
$Fe_{s} + Ga_{i}$  IC to be lower in energy.
On the other hand, for $1.5$ and $2$ ML of Fe  we calculate the
$Fe_{i}$ IC to be lower in energy. It is the exchange energy that
stabilizes the $Fe_{i}$ configuration, since in a non-spin polarized calculation we
find again $Fe_{s} + Ga_{i}$  to be lower. (This will be
further discussed in the next section.) 
If both As and Ga segregation are included (section VI), we find Fe to
dissolve GaAs independent of thickness.

\begin{table}
\caption{Calculated $c/<a>$ ratio and in-plane lattice contraction of
the unit cell along the $<1\bar{1}0>$ and $<110>$ direction for the lowest
energy configuration of the specified Fe film.} 
\begin{tabular}{|l|l|l|l|}
\hline \hline
                            & $c/<a>$   & $<1\bar{1}0>$  & $<110>$ \\
\hline

0.5 ML Fe                            &  1.07  & -1.15 \% & -6.92 \%   \\
1 ML Fe + 0.5 ML As                            &  1.03  & -0.51 \% &  -3.66 \%   \\
1 ML Fe + 1ML As                            &  1.02      & +0.86 \% &  -2.79 \%  \\
2 ML Fe + 0.5 ML As                            &  1.10      & -0.81 \% &  -5.61 \%  \\
2 ML Fe + 1 ML As                           &  1.05       & +0.86 \% &  -2.79 \%   \\
5 ML Fe + 0.5 ML As                       & 1.04  & -1.15 \% & -2.74 \% \\
5 ML Fe + 1 ML As                       & 1.03  & +0.51 \% & -1.83 \% \\
\hline \hline
\end{tabular}
\end{table}

In a recent x-ray absorption fine structure spectroscopy (XAFS) study
Gordon et al \cite{gordon} found the Fe film on GaAs(100) to be tetragonal distorted
relative to bulk bcc Fe. The measured distortion involved an in-plane
contraction and an expansion perpendicular to the GaAs surface to give a
$c/a$ ratio of $1.03$. In tab.1 we have collected the calculated $c/a$
ratio as a function of the Fe film thickness and As coverage. 
Our results agree rather good with experiment.
We find in agreement with experiment for our energetically lowest
interface configurations an in-plane contraction and an expansion
perpendicular to the GaAs surface.
In addition we find a rather large asymmetry in the in-plane
contraction. In general the contraction along the $<110>$ direction is
larger than 
along the $<1\bar{1}0>$ direction. 
We even find an expansion along the $<1\bar{1}0>$ direction for all studied Fe coverages
which are covered by a complete As monolayer. 

The expansion perpendicular to the GaAs surface is connected
to the Ga-content within the Fe-film, because for our calculations without 
any Ga
atoms within the Fe-film we find instead a contraction perpendicular to the GaAs
surface. The in-plane asymmetry is caused by the directional bonds of
the GaAs substrate. On the As-terminated surface, the
As dangling bonds are oriented along the $<1\bar{1}0>$ direction. 
Because of the Fe-As interaction and the $Fe_{s}+Ga_{i}$ IC, an
expansion/contraction along the $<1\bar{1}0>$  direction has to
optimize the Fe-As interaction. In contrast, an expansion/contraction
along the $<110>$ direction is a reaction on the expansion/contraction
along the other two directions in order  
to optimize the overall volume. Therefore, the
$<1\bar{1}0>$ and $<110>$ direction are asymmetric. For a Ga-terminated
surface the same is valid, but the $<110>$ and $<1\bar{1}0>$ directions
are interchanged. Note, that this in-plane asymmetry explains the
in-plane uniaxial magnetocrystaline anisotropy \cite{brockmann,sjostedt}.

\section{Magnetic Properties}

\begin{figure}
\resizebox{6.5cm}{!}{\includegraphics[angle=270]{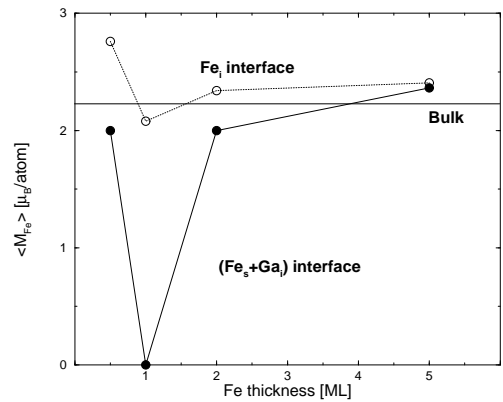}}
\caption{Average Fe magnetization,$<M_{Fe}>$, as a function of the Fe film thickness for two
interface configurations. Solid
(dotted)  line corresponds to the $Fe_{s}+Ga_{i}$ ($Fe_{i}$) IC.
The solid horizontal line indicates the calculated Fe bulk magnetic moment.}
\end{figure}

\begin{table}
\caption{Total magnetic moment and atomic distances for the respective closest topmost
atoms for 1ML of Fe for the three interface configurations shown in Fig.6.
Ga-As (h) (GaAs (v)) indicates the horizontal (vertical) distance.}
\begin{tabular}{|l|l|l|l|}
\hline \hline
                   & Fe$_{i}$ &
Fe$_{s}$+Ga$_{i}$ & Fe$_{s}$+Ga$_{i}$           \\
                            &    & start  & final \\
\hline

Fe - As                            &  2.61 $\AA$  & 2.42 $\AA$ &  2.32
$\AA$ \\
Fe - Ga                            &  2.96 $\AA$  & 2.42 $\AA$ &  2.51 $\AA$   \\
Ga - As (h)                            &  -      & 2.87 $\AA$ &  2.82 $\AA$  \\
Ga - As (v)                            &  -      & 2.42 $\AA$ &  3.61 $\AA$  \\
Ga - Ga                           &  -       & 2.44 $\AA$ &  2.44 $\AA$   \\
$ M_{cell}$                       & 4.63 $\mu_{B}$ & - & 0.01 $\mu_{B}$ \\ 
\hline \hline
\end{tabular}
\end{table}

In Fig.5 we show the average Fe magnetic moment as a function of the Fe
thickness for the $Fe_{i}$ and $Fe_{s} + Ga_{i}$ IC. 
Surprisingly, we calculate for the thickness of $1$ ML of $Fe$ 
the Fe magnetic moment to be zero. In Fig.6 we show the atomic

\begin{figure}
\resizebox{6.5cm}{!}{\includegraphics[angle=270]{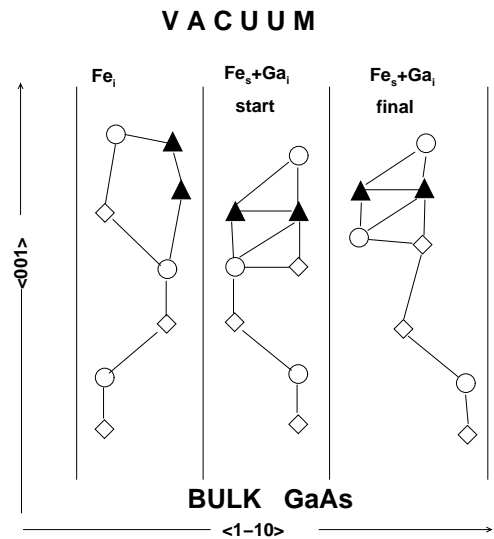}}
\caption{Geometrical structure of 1 ML Fe on top of GaAs(100). Circles
(diamonds)
represent As (Ga) and triangles represent Fe. The three columns
correspond to three different configurations as discussed in the text.}
\end{figure}

configuration for $Fe_{i}$ and $Fe_{s} + Ga_{i}$ for the Fe thickness of
1 ML. It can be seen by comparing the
starting and final positions, that for $Fe_{s} + Ga_{i}$ the Fe-As bond
distance has decreased (tab.2). In addition the 'vertical' Ga-As bonds 
are broken and the three top layers became shifted along the $<1\bar{1}0>$ direction.
This suggests
that the three top layers now form a separate 2-dimensional phase. 

\begin{figure}
\resizebox{6.5cm}{!}{\includegraphics[angle=270]{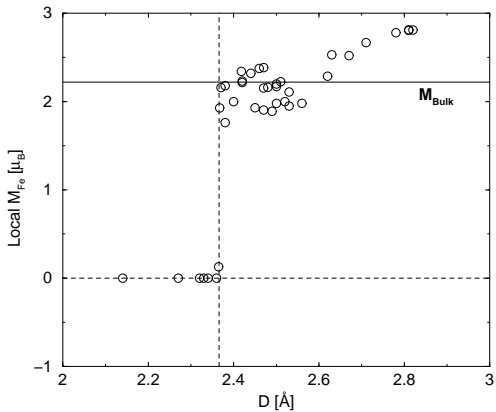}}
\caption{Local magnetic moment of Fe, $M_{Fe}$, as a function of the Fe-As distance
, $D$. Solid
(dashed)
horizontal line indicates the bulk Fe magnetic moment (zero-line) and
the dashed vertical line at $2.37$ \AA\ indicates the Fe-As distance at
which the Fe magnetic moment becomes zero.}
\end{figure}

In order to understand
the decreased magnetic moment, we plot in Fig.7 the Fe magnetic moment
as a function of the Fe-anion distance , $D$, (i.e. the Fe-As distance) for all different configurations that
we have calculated. It can be seen that for $ D= 2.36 $
\AA , the Fe magnetic moment disappears. We interpret this to be caused
by 
$pd$-hybridization: The Fe spin-polarization is driven by
the on-site exchange interaction. Therefore the Fe magnetic
moment decreases with increasing delocalization of    
the Fe $d$-states. Because, the delocalization of the Fe $d$-states increases
with increased overlap between the Fe $d$-states 
and As $p$-states, we find that the Fe magnetic moment is quenched for
small Fe-As bond distances. 
The spread in magnetic moments for bond lengths above $2.36$ \AA~ is due
to the other factors influencing the Fe magnetic moment, like number of
Fe neighbours and number of vacant neighbours.  

\begin{figure}
\resizebox{6.5cm}{!}{\includegraphics[angle=270]{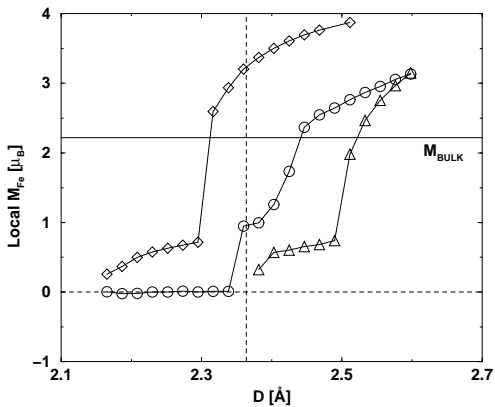}}
\caption{Bulk magnetic moment per Fe atom for zincblende FeAs (circles),
zincblende FeSe (diamonds), and zincblende FeTe (triangles) as a function of
the Fe-anion distance, $D$. The dashed vertical line is the same as in
Fig. 7 shown here for comparison.}
\end{figure}

Our interpretation is supported by the behaviour of the Fe magnetic
moment in the bulk FeAs zincblend compound (circles in Fig.8). We show the Fe
magnetic moment as a function of the Fe-anion distance ($D$) in the zincblend
lattice. The same trend as before is observed, namely the Fe magnetic moment
becomes smaller at a critical $D$.  
In contrast to Fe on GaAs, the Fe magnetic moment is not
quenched on ZnSe \cite{sanyal}. In order to understand the difference
between GaAs and ZnSe we also calculated zincblende FeSe and for
comparison zincblende FeTe.
In all three cases, i.e. FeAs, FeSe, and FeTe, the Fe $d$-electrons
hybridize with the respective anion's $p$-electrons.
Therefore there exists a bond distance at which the Fe magnetic moment
becomes quenched (Fig.8). The exact value of the critical $D$ depends on the
system in question. For a given lattice constant (set by for example the
semiconductor host), the anion will occupy a certain fraction of the
unit cell volume that depends on its atomic radius. GaAs and ZnSe have
about the same lattice constant. The As atomic radius (1.14 \AA ) is
larger than the Se radius (1.03 \AA ). The $pd$-overlap between Fe and Se
is therefore (for the same lattice constant) smaller than between Fe and As
and the Fe magnetic moment starts to become quenched for a smaller $D$.
On the other hand the Te radius (1.23 \AA ) is larger and the
magnetic moment becomes quenched already for a larger $D$. 

As clearly visible from Fig.8 there exist high-spin,
low spin, and zero moment phases as a function of $D$. The energy
difference between these phases becomes rather small for a given $D$.
This will be further investigated in the future. We conclude that
the magnetic configuration of Fe on GaAs will be determined by
$pd$-hybridization, but that the specific magnetic phase of the Fe-As
complex depends on small variations of the Fe-As bonding.
For example, in earlier publications \cite{erwin,hong} an
antiferromagnetic (AFM) solution of
 the Fe film has
been discussed and calculated. Especially for the $1$ ML thick Fe film an AFM
solution was found. We also investigated the possibility of an AFM
solution. For a $1$ ML Fe film covered with a complete ML of As we find
an AFM solution for a $Fe_{s}$ interface configuration ($0.22 \mu_{B}$,
$ -0.22 \mu_{B}$). The involved magnetic moments are so small because of
the Fe-As $pd$-hybridization as discussed above. The total energy is $3$
meV higher than a solution with a zero magnetic moment on both Fe sites,
i.e the AFM solution is almost degenerate with the zero moment per Fe
atom solution. For the $Fe_{s} + Ga_{i}$ interface configuration we
again find an AFM solution ($0.05 \mu_{B}$, -$0.05 \mu_{B}$) being $10$ meV higher than a solution having
a total magnetic moment of $0.5 \mu_{B}$. 
Another example we find from our calculations is a ferrimagnetic solution. The structure consists of a
$5$ ML thick Fe film on top of GaAs covered with $0.5$ ML
As and an additional $0.5$ ML As
within the Fe film. The top Fe atom has a magnetic moment of
$-2.1 \mu_{B}$, whereas the two Fe atoms/cell in the next layer beneath
have a
magnetic moment of $0.95
\mu_{B}$ and $1.2 \mu_{B}$, respectively. Below this layer the
additional
$0.5$ ML As is located. The rest of the Fe atoms have a magnetic moment
close
to or larger than the
bulk value. The average magnetic moment for this ferrimagnetic solution
is $1.55 \mu_{B}$ per Fe atom.

\begin{figure}
\resizebox{6.5cm}{!}{\includegraphics[angle=270]{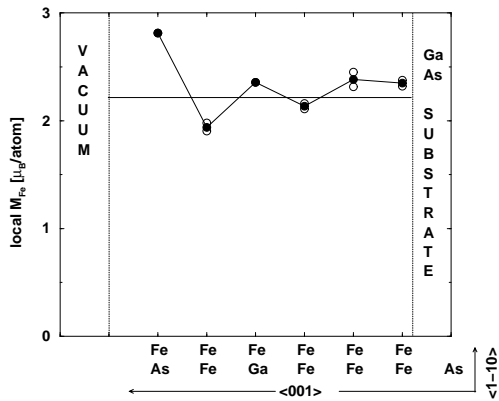}}
\caption{Local Fe magnetic moment, $M_{Fe}$, profile for 6ML Fe ( 10 Fe atoms/cell) on top of GaAs covered with 0.5 ML
As and 0.5 ML Ga within the Fe film. The horizontal line indicates the
calculated bulk Fe magnetic moment. The filled (open) circles correspond to the
average (individual) Fe magnetic moment of the respective layer.
On the x-axis the atomic structure is sketched along the $<001>$ direction
projected on the $<1\bar{1}0>$ direction.}
\end{figure}
\begin{figure}
\resizebox{6.5cm}{!}{\includegraphics[angle=270]{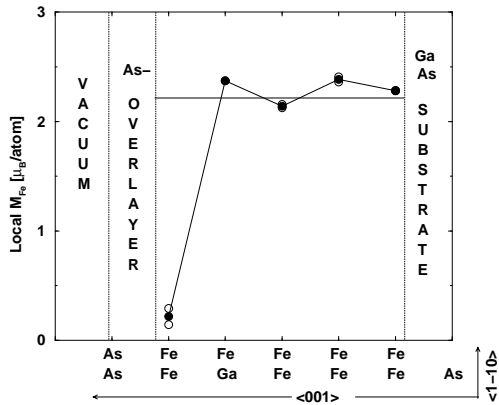}}
\caption{Same as Fig.9, but for 5ML Fe (9 Fe atoms/cell) covered with 1ML As.}
\end{figure}

Since the magnetic phase diagram of the Fe-As interaction is rather
complex, it very much depends on small details of the Fe film
configuration which magnetic moment will be measured.
But the physical mechanism behind the reduction of the Fe magnetic
moment on top of GaAs is without any doubt the Fe-As
$pd$-hybridization.

As can
be seen from Fig.5, the Fe magnetic moment is larger for $Fe_{i}$  than
for $Fe_{s} + Ga_{i}$.
In general the
$Fe_{s} + Ga_{i}$ configuration has a smaller Fe-As bond distance than $Fe_{i}$, which
explains the smaller magnetic moment for the $Fe_{s} + Ga_{i}$
configuration. For $1.5$ and $2$ ML of $Fe$, the
energy gain due to increased Fe-As binding is smaller than the energy
gain due to the increased spin-polarization of the Fe atoms. This
explains why $Fe_{i}$ is lower than $Fe_{s} + Ga_{i}$ for $1.5$ and $2$
ML of $Fe$ (see discussion above and Fig.4) 

In Figs.9 and 10 we show the structure and magnetic profile of a 5ML
Fe film on top of GaAs(100) with one Ga atom within the Fe film for two
segregation profiles: In Fig.9
for $0.5$ ML As (i.e one As surface atom per cell) on top of the Fe film;
in Fig.10 for 1 ML As (i.e  two As atoms per cell) on top of the Fe film.
The Ga atom does not influence the Fe magnetic moment, i.e the Ga-Fe
interaction is very weak. On the other hand, the Fe-As interaction
quenches the magnetic moment (as discussed before). For the 1ML As
coverage, the Fe magnetic moment of the top Fe layer is thus almost zero
(Fig.10).
If Fe is covered with only 0.5 ML As, the Fe-Fe interaction is stronger
than the Fe-As interaction and the Fe magnetic moment is not reduced (Fig.9).

From our calculations we can conclude:
\begin{equation}
E_{Fe}^{xc} > E_{Fe-As}.
\end{equation}
\begin{equation}
E_{Fe-As} > E_{Fe}^{xc} + E_{Fe-Fe}.
\end{equation}
\begin{equation}
E_{Fe}^{xc} + E_{Fe-Fe} > E_{As-Fe-As}.
\end{equation}
Here $E_{Fe-As}$ is the Fe-As surface bond formation energy, $E_{Fe}^{xc}$ is the energy
gain due to the spin-polarization of the Fe atom, $E_{Fe-Fe}$ is the
Fe-Fe surface bond formation energy, and $E_{As-Fe-As}$ is the bond formation
energy between a Fe surface atom
bonded to two As atoms.
These three equations explain the calculated magnetic properties.\\ 
Equation 3: \\ 
This equation states that the Fe magnetic moment will be quenched if Fe
only has one or more As neighbours. 
For example, the Fe magnetic moment becomes zero for 0.5 ML Fe with a $Fe_{s}$ IC
(Fig.3). Here the Fe atom has no other Fe around it but an As atom.\\ 
Equation 4: \\ 
This equation states that the Fe magnetic moment is not
quenched if the Fe atom is close to one As atom and at least one Fe atom.
This is the case for the $Fe_{i}$ IC of 1ML Fe on top of GaAs (Fig.6 and
tab.2).\\
Equation 5: \\
The Fe magnetic moment becomes quenched again, if Fe is bonded to two As
atoms (Fig.10, Fig.6), independent on the number of Fe neighbours.

\begin{figure}
\resizebox{6.5cm}{!}{\includegraphics[angle=270]{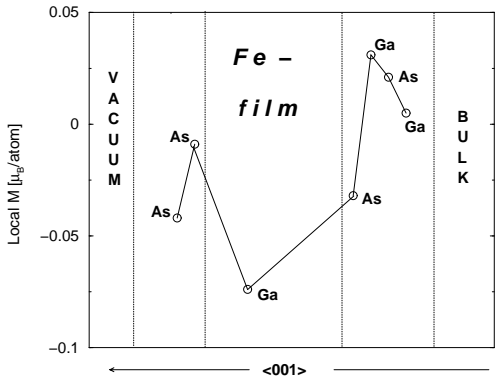}}
\caption{Local magnetic moment,$M$, profile of Ga and As for the geometrical structure of
5 ML Fe on top of GaAs covered with 1 ML As and 0.5 ML Ga within the
Fe film. The Fe magnetic moment is not shown here.}
\end{figure}

For completeness, we show in Fig.11 the spin-polarization of the GaAs host. The Fe magnetic
moment is not shown here (see Fig.10). The induced spin-polarization is
mostly antiparallel and rather small. 

In a recent X-ray absorption study \cite{freeland}, the number of Fe $3d$-holes was determined as a function of the Fe film thickness on $n$-doped GaAs. Freeland et
al find the number of holes to increase with decreasing Fe film
thickness. They explain this due to charge transfer from Fe to As
which they also believe to be the cause for the reduced Fe magnetic
moment.

\begin{figure}
\resizebox{6.5cm}{!}{\includegraphics[angle=270]{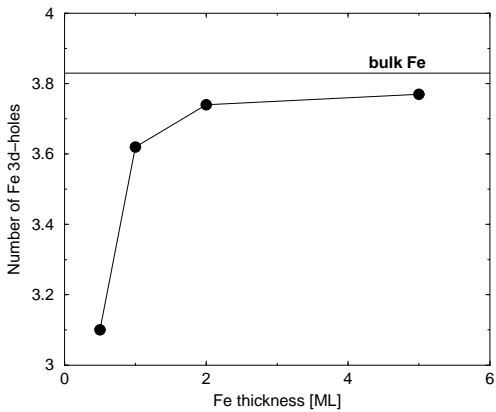}}
\caption{Calculated number of Fe $3d$-holes as a function of the Fe film thickness.
The horizontal line indicates the number of Fe $3d$-holes for Fe bulk.}
\end{figure}

Our calculation describes the charge transfer between Fe and $p$-doped GaAs,
because due to numerical reasons the Fermi energy of bulk GaAs is fixed 
at the top of the valence band edge.
From our calculations we directly get the number of Fe $3d$-holes
as a function of the Fe film thickness (Fig.12).
Our absoult number of
the Fe $3d$-holes depends of course on the chosen
Wigner-Seitz radius for Fe, but the trend we find to be
independent on the chosen Wigner-Seitz radius. In contrast to
experiment we find an increase of the holes with increasing Fe film
thickness. 

Our explanation is the following: The Fe-As $pd$-hybridization
determines the Fe magnetic moment, not a charge transfer between Fe and
As. The electron transfer between Fe and GaAs is different for $n$-type
and $p$-type conditions, because for $n$-type ($p$-type) GaAs, the Fermi
level of Fe 
has to align with the conduction (valence) band of GaAs. 
Under $p$-type conditions,   
a consequence of the hybridization is that on average there are
slightly more delocalized electrons on the Fe site, i.e. the number of $3d$-holes
decreases. The thicker the Fe film becomes, the more Fe atoms are without an As neighbour
, which is why the number of $d$-electrons (holes) decreases (increases) with 
increasing Fe thickness. Under $n$-type conditions it is plausible to
assume vice versa that on average there are slightly less delocalized electrons on
the Fe site.
We therefore predict that the experimentally observed charge transfer differs between
$p$-doped GaAs and $n$-doped GaAs and that similar experiments performed
instead on $p$-doped GaAs should
find an increase of the $3d$-holes with increasing thickness in agreement
with our calculations.

\section{Surface Segregation}
It is well known that at metal-semiconductor interfaces the
semiconductor constitutents segregate towards the surface. For example,
for Fe on GaAs it is known that As segregates to the surface, whereas
Ga is not found at the surface but within the metal film
\cite{chambers,ruckman,kneedler}. It is found that the segregation of As is independent
of temperature, but the segregation of Ga is dependent on temperature.
A quantitative understanding of segregation is obtained with a simple
model put forward by Weaver et al. \cite{weaver}. The free energy of
segregation is determined by the strain energy and surface energy, the
latter of which
they estimated with the cohesive energy.   
There model predicts for Fe on GaAs both Ga and As to segregate to the
surface,
whereas in experiment only As surface segregation is observed.
In the following we resolve this discrepancy.  

\begin{figure}
\resizebox{6.5cm}{!}{\includegraphics[angle=270]{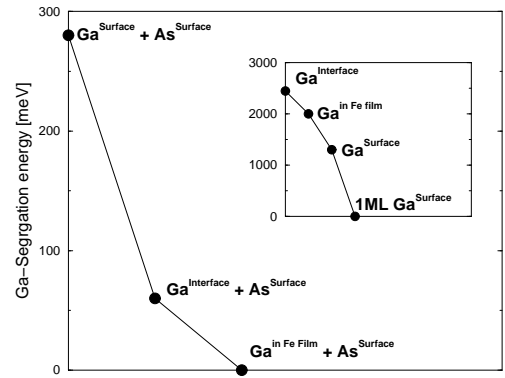}}
\caption{Ga segregation energy for four different Ga configurations as discussed
in the text for 5ML Fe on top of GaAs.}
\end{figure}
\begin{figure}
\resizebox{6.5cm}{!}{\includegraphics[angle=270]{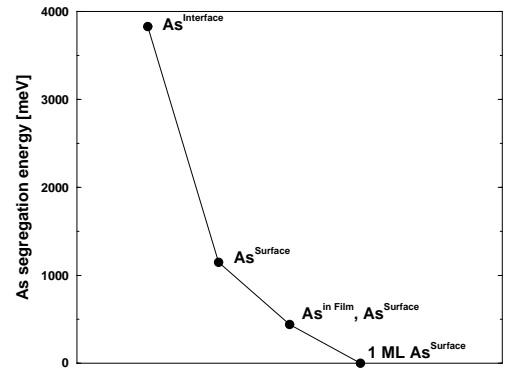}}
\caption{As segregation energy for four different As configurations as discussed
in the text for 5ML Fe on top of GaAs.}
\end{figure}

In Figs.13 and 14, we show the Ga (respective As) segregation energy
relative to the energetically lowest configuration. For Ga we have to
consider two cases separately: (i) no As has segregated to the surface
and (ii) As has segregated to the surface. In
case (i) (insert of Fig.13), Ga prefers to segregate to the surface (1ML
Ga$^{Surface}$). The segregation energy of 1ML Ga amounts to 2.4 eV. 
In case (ii) where As already has
segregated to the surface (Fig.13), Ga prefers to leave
the interface (Ga$^{Interface}$) and stay somewhere within the Fe film 
(Ga$^{in Fe-film}$). It is more costly (280 meV) for the Ga to be in the surface layer
together with an As atom (Ga$^{Surface}$). Therefore, if As has
segregated to the surface, no Ga will be found at the surface.

The As segregation energy (Fig.14) shows the same trend as the Ga
segregation energy (inset of Fig.13), but the segregation energy of one As atom
to the surface is 1.5 eV larger than for one Ga atom.  
Notice, that both Ga and As prefer to be within the Fe film rather than at the
interface.
In summary, this suggests the following: 
On top of the Fe film always As will be found
independent of the GaAs surface termination.
The segregation profiles should more or less be independent on the
GaAs surface termination, because 
the Fe-As interaction is much stronger than the Fe-Ga interaction. 

The  As segregation path is a result of the
lattice relaxation due to the chemical interaction between the Fe and
As, which implies that As segregates already at $T = 0$ K.
The segregation is thus {\bf not} diffusion controlled, but only
controlled by chemical bonding, which is in agreement with
experiment.
We find Ga to leave the interface, but only to segregate to the
surface, if no As already has segregated. In contrast to As, the segregation
of Ga, does not take place at $T=0$ K. There exists an activation barrier for the
segregation, i.e. the segregation is diffusion controlled. The amount of Ga on top of or within the Fe film is therefore
strongly dependent on temperature, whereas the amount of As on top of 
the Fe film is only weakly dependent on temperature. This is in
agreement with an experimental study of the Fe/GaAs interface
interdiffusion \cite{rahmoune}.

Regarding the Weaver model, we find that Ga segregates to the surface
(in agreement with the Weaver model), if no As has segregated. 
If As has segregated, Ga will stay
within the Fe-film ( in agreement with experiment). 

\section{Discussion and Summary}
We find As to segregate to the surface on top of the Fe film independent
of termination, which is in agreement with experiments. The Fe/GaAs
interface is not stable against further segregation. An As atom within
the Fe film has a lower energy than at the interface. In an
experiment one will thus always find As within the Fe film, where the
Fe-As $pd$-hybridization will quench the Fe magnetic moment. Since Ga also
has a lower energy within the Fe film than at the interface, one will
also always find Ga within the Fe film. Ga itself does not influence the
magnetic moment (Fig.9), but probably Ga in the Fe film will prevent
further As segregation. This would explain the much lower thickness of
the magnetically quenched Fe layers for Ga terminated samples.
The thickness of the magnetically quenched Fe layers is then determined
by the probability of finding As and Ga within the Fe film. This
probability depends on the termination and the Fe growth conditions.
Our results presented here for Fe on GaAs are most likely also valid for
other semiconductor substrates. For example, we find more or less the
same structures for Fe on ZnSe \cite{sanyal} and the segregation behaviour
is the same, but the Fe-Se hybridization is much weaker.

\acknowledgements
We are grateful to the Swedish foundation for strategic research (SFS), the Swedish Research Council (VR), the G\"{o}ran Gustafsson
foundation, and the European RTN network on "computational spintronics" for 
financial support. We thank the Swedish supercomputer center
(SNAC) for providing us with supercomputer time.

\end{document}